\documentclass[%
 reprint,
nofootinbib,
 amsmath,amssymb,
 aps,
 prd,
]{revtex4-1}

\usepackage{graphicx}
\usepackage{dcolumn}
\usepackage{bm}
\usepackage{microtype}
\numberwithin{equation}{section} 
\usepackage{hyperref}
\usepackage{xcolor}

\begin{document}


\title{Kinetically coupled dark energy}

\author{Bruno J. Barros}
\affiliation{Instituto de Astrof\'isica e Ci\^encias do Espa\c{c}o,\\ 
Faculdade de Ci\^encias da Universidade de Lisboa,  \\ Campo Grande, PT1749-016 
Lisboa, Portugal}%

\date{\today}

\begin{abstract}

The main goal of this work is to propose a generalized model of interacting dark energy which allows for the kinetic term of a scalar field to couple to the matter species {\it a priori} in the action. We derive the modified field equations, and present novel cosmological solutions for a specific coupled model. One alluring consequence is the emergence of solutions allowing for an early scaling regime, possible due to two novel critical points, followed by a period of accelerated expansion. Using a dynamical system analysis, we show that the presence of the coupling may alter the dynamical nature of the critical points and can be used to enlarge the existence and stability regions of these. Using constraints from Planck data we are able to find an upper bound on the coupling parameter. Finally, it is shown how this theory encapsulates a wide variety of dark energy models already present in the literature. 
\end{abstract}

\maketitle


\section{Introduction}

The energy source driving the late-time acceleration of the Universe \cite{Perlmutter:1998np,Riess:1998cb} remains unknown to this day, and has been baptized as {\it dark energy} (DE). The simplest candidate for this substance is the cosmological constant $\Lambda$ \cite{Peebles:2002gy} which bears a fundamental role in the standard model of Cosmology, the $\Lambda$CDM model. However, this cosmological constant paradigm faces some unanswered problems \cite{Peebles:2002gy,Weinberg:2000yb} and so, cosmologists explore other possibilities for this dark species. This quest usually focuses on one of the following two alternatives: either one assumes a modification of the gravitational sector, e.g. modified gravity theories \cite{Nojiri:2006ri,Sotiriou:2008rp,DeFelice:2010aj,Clifton:2011jh}; or one introduces a fluid with an exotic character, such as scalar fields \cite{Copeland:1997et,Chiba:1999ka}. This work will focus on the latter. When portraying dark energy as a cosmic fluid with the aid of canonical scalar fields, it is usual to refer to these theories as quintessence models. These were first proposed in \cite{Wetterich:1994bg} and one of their enticing consequences is the emergence of scaling solutions \cite{Copeland:1997et,Albuquerque:2018ymr} which might be used to alleviate the cosmic coincidence problem \cite{Zlatev:1998tr,Chimento:2003iea}.

It is only natural to assume that the quintessence field interacts with other matter sources. Coupled models were then proposed in \cite{Wetterich:1994bg,Amendola:1999er} with particular viable scaling solutions \cite{Barreiro:1999zs}. Certain types of couplings can be generated in scalar-tensor theories \cite{PhysRevLett.64.123,Amendola:1999qq,Pettorino:2008ez}, by doing a conformal transformation. In these conformally coupled theories, the matter fields are assumed to experience a different metric, $\tilde{g}_{\mu\nu}$, related to the one describing the gravitational sector, $g_{\mu\nu}$. In such case, the matter Lagrangian couples to a field dependent function in the Einstein frame. The generalization for multiple scalar fields and several matter fluids was studied in \cite{Amendola:2014kwa}, where it was found that, in the case of exponential potentials, the scaling solutions can be described in terms of a single field. A study on the linear perturbations was also done in \cite{Amendola:2003wa,Amendola:2002mp} and the influence of the coupling on structure formation and halo mass functions was investigated in \cite{Maccio:2003yk,Tarrant:2011qe}. The evolution of the matter fluctuations in such coupled models are modified by the emergence of a fifth force, induced by the scalar degree of freedom, and an additional damping term. The interplay between these two effects has an impact on the growth rate of the perturbations and has been used \cite{Barros:2018efl}, for example, to alleviate the tension on the perturbation parameter $f\sigma_8$ observed for the $\Lambda$CDM model \cite{Battye:2014qga,Macaulay:2013swa}.

In the literature, a large number of theories regarding dark energy either impose the couplings at the level of the field equations \cite{Leithes:2016xyh,Piloyan:2014gta,Tsujikawa:2010sc,Cai:2004dk,Zimdahl:2001ar} or they arise through a scalar-tensor type theory \cite{Amendola:1999qq}. A structured Lagrangian formalism for coupled quintessence was examined in \cite{Koivisto:2005nr,Bean:2000zm} where the matter species were coupled to a function solely of the field. In \cite{Farrar:2003uw}, a coupled model where the mass of the dark matter particles varies with the value of the dark energy $\phi$ field through a Yukawa coupling was explored. The action formalism for the case where the dark matter fields are described in terms of either a wave function $\psi$ for $1/2$-spin particles, or scalar dark matter particles, was presented. The stability of quintessence models regarding quantum fluctuations was investigated in \cite{Doran:2002bc} for both interacting (considering couplings to fermions) and noninteracting models. 
Couplings to matter within Horndeski theories were explored in \cite{Gomes:2015dhl}, where the form of the Lagrangian which allows for scaling solutions to exist was derived .

In nature, it seems like most species naturally interact with one another, so it is not unreasonable to consider kinetic terms dynamically influencing other components. Some models that explore this relation have been proposed. Mimetic models \cite{Chamseddine:2013kea,Sebastiani:2016ras} with couplings incorporating the field derivatives have been studied in \cite{Shen:2017rya,Vagnozzi:2017ilo}. In these theories, the derivatives of the mimetic field, portraying dark matter, couple linearly to the matter current in the action. It was shown that, in such cases, by assuming a shift symmetry, the mimetic field can only present derivative couplings (not directly involving the field itself) to the matter content of the Universe. The inclusion of derivative interactions, at the level of the action, within quintessence models, was mentioned in \cite{Pourtsidou:2013nha}, where linear couplings of the field's derivative to the fluid's four-velocity $u_{\mu}$ were thoroughly explored, considering an object of the form $u^{\mu}\nabla_{\mu}\phi$. These so-called scalar-fluid models are written using Brown's formalism \cite{Brown:1992kc} and consider derivative interactions expressed as a linear coupling of the field derivative to the fluid vector-density particle-number flux $J^{\mu}$ (related with the fluid's four velocity $u^{\mu}$) \cite{Boehmer:2015sha,Boehmer:2015kta}. A dynamical system analysis was done in \cite{Dutta:2017kch} and the analysis of the perturbations and consequent formation of large scale structures was investigated in \cite{Koivisto:2015qua}. Nonminimal kinetic couplings to curvature were explored in \cite{Granda:2009fh,Granda:2011eh}. One alluring consequence found was the existence of late time solutions leading to accelerated expansion. A final example of models in which the field's velocity has an intimate relation with the matter sector are the so-called disformally coupled theories \cite{Bekenstein:1992pj,Zumalacarregui:2012us,Zumalacarregui:2010wj,vandeBruck:2015ida}. These models generalize the notion of conformal transformations by allowing the rescaling of the metric to take into account the kinetic term of the scalar field, through a disformal transformation, for example of the form $\tilde{g}_{\mu\nu} = C(\phi)g_{\mu\nu} + D(\phi)\partial_{\mu}\phi\partial_{\nu}\phi$ \cite{vandeBruck:2016jgg}. The rescaled metric, in which the matter fields live, is now intimately connected with the field's velocity when $D(\phi)\neq 0$. The analysis of spherical collapse and cluster number counts in disformally coupled theories was explored in \cite{Sapa:2018jja}. A generalization of derivative couplings in scalar-fluid models can be found on the last pages of \cite{Koivisto:2015qua} with the aid of disformal couplings.

In this work we propose a generalized form for interacting dark energy models by allowing a general scalar field $\phi$, with Lagrangian density $P(\phi,X)$, where $X \equiv -\frac{1}{2}\partial^{\mu}\phi \partial_{\mu}\phi$, to kinetically couple to the matter sources. That is, we will assume that the kinetic term $X$ of the scalar field can directly couple to matter fields at the level of the action. We will assume that this interaction is described through a general function expressed in terms of the field and its derivatives, $f(\phi,X)$, in the Lagrangian. This term takes the form $f(\phi,X)\tilde{\mathcal{L}}_m$, where $\tilde{\mathcal{L}}_m$ is the matter Lagrangian. After deriving the cosmological field equations for the proposed theory, we solve them for a particular model and show that the presence of the coupling allows for solutions with an early scaling regime followed by an accelerated expansion period, when this $f$ function depends solely on the kinetic term. We will show that the scaling regime, useful to alleviate the cosmic coincidence problem, is only possible due to the emergence of two new critical points when the coupling is present. By rewriting the equations of motion as a first order dynamical system, we study the nature of the kinetic coupling on the overall evolution of the cosmological parameters. We find an upper bound for the coupling parameter using constraints from Planck data. Finally, we show how specific cases of this theory reproduce a large number of dark energy models already explored in the literature.

This work presents the following structure: the general equations of the proposed theory are derived in \ref{modell} and in subsection \ref{specificcases} we show how already known models of dark energy can naturally emerge as special cases of the theory. In \ref{ex1} we demonstrate a particular example, where we specify the Lagrangian for the species, the coupling function and follow to rewrite the equations of motion as a dynamical system. The study regarding the nature of the critical points of the linear autonomous system is presented in \ref{tabelaP}. We numerically solve the equations in \ref{dsanalysis}, present the solutions and analyse them. Finally we conclude in section \ref{conclusions}.

\section{Model}
\label{model}

\subsection{Action and field equations}
\label{modell}

The geometry of our Cosmology can be described by a smooth differentiable manifold $\mathcal{M}$ endowed with a metric {\bf g} which provides information regarding distance measurements on the Universe. The pair $(\mathcal{M},\,${\bf g}$)$ forms a smooth Riemannian space whereupon we will allow the matter species to live.

Regarding the matter sector we will consider a two-component Universe, consisting of one scalar field $\phi$, coupled to one matter fluid, with Lagrangian densities $\mathcal{L}_{\phi}$ and $\tilde{\mathcal{L}}_{m}$ respectively. The total action, minimally coupled to Einstein gravity, of the proposed kinetically coupled model can be written as
\begin{equation}
\label{action}
\mathcal{S} = \int d^4 x \sqrt{-g} \left[ \frac{R}{16\pi G} + P(\phi,X) + {f}(\phi,X)\tilde{\mathcal{L}}_m(g_{\mu\nu},\psi) \right],
\end{equation}
where $g\equiv {\rm det}\,g_{\mu\nu}$, being $g_{\mu\nu}$ the components of the metric tensor, $R$ is the Ricci scalar portraying the gravitational sector, constructed from the metric $g_{\mu\nu}$, $\psi$ represents the matter field, and the function $f(\phi,X)$ entails the information on how the $\phi$ field couples to the matter species. Note that the novelty of this work is to allow the function $f$ to depend also on the kinetic term of the field, $X \equiv -\frac{1}{2}g^{\mu\nu} \partial_{\mu}\phi \partial_{\nu}\phi$.
From Eq.\eqref{action} we can identify the scalar field Lagrangian as $\mathcal{L}_{\phi} = P(\phi,X)$. 

Varying the action Eq.\eqref{action} with respect to the metric holds the modified field equations,
\begin{equation}
\label{EE}
\frac{1}{8\pi G} G_{\mu\nu} = T^{(\phi)}_{\mu\nu} + {f}\, \tilde{T}^{(m)}_{\mu\nu} + {f}_{,X} \tilde{\mathcal{L}}_m \partial_{\mu}\phi \partial_{\nu}\phi,
\end{equation}
where
\begin{equation}
T^{(i)}_{\mu\nu} \equiv  -2\frac{\delta \mathcal{L}_{i}}{\delta g^{\mu\nu}} + \mathcal{L}_{i}g_{\mu\nu}
\end{equation}
are the energy-momentum tensors of the $i$th species and $G_{\mu\nu}$ are the components of the Einstein tensor. In order to avoid extensive expressions we simply write $P(\phi,X)\equiv P$ and ${f}(\phi,X)\equiv {f}$. The fact that we are considering couplings that also depend on the kinetic term of the scalar field, gives rise to a new interaction term - the last one on the right-hand side of Eq.~\eqref{EE} - in the Einstein equations. As we will  see, this will alter the conservations relations, and thus have an impact on the overall dynamical evolution of the system.

We can rewrite the field equations in a more familiar form, by defining
\begin{equation}
\label{flm}
\mathcal{L}_m (g_{\mu\nu},\psi,\phi,X)\equiv f(\phi,X) \tilde{\mathcal{L}}_m(g_{\mu\nu},\psi).
\end{equation}
We will work with this quantity, representing an effective matter Lagrangian encapsulating the effect of the coupling on the fluid {\it per se}, which is more convenient and simplifies both the analysis and the equations. In this framework, the stress-energy tensors are related through,
\begin{eqnarray}
T^{(m)}_{\mu\nu} &=& -2\frac{\delta \mathcal{L}_{m}}{\delta g^{\mu\nu}} + \mathcal{L}_{m}g_{\mu\nu} \nonumber \\ 
&=& {f}\,\tilde{T}^{(m)}_{\mu\nu} -2 \tilde{\mathcal{L}}_m  \frac{\delta {f}}{\delta g^{\mu\nu}}\nonumber\\
&=& {f}\, \tilde{T}^{(m)}_{\mu\nu} + {f}_{,X} \tilde{\mathcal{L}}_m \partial_{\mu}\phi \partial_{\nu}\phi,
\end{eqnarray}
and we can rewrite the field equations, Eq.\eqref{EE}, more conveniently as
\begin{equation}
\label{EEs}
\frac{1}{8\pi G}G_{\mu\nu} = T^{(\phi)}_{\mu\nu} + T^{(m)}_{\mu\nu}.
\end{equation}

The action Eq.~\eqref{action} leads to the following equations of motion for $\phi$,
\begin{gather}
\label{motion}
P_{,\phi}+P_{,X}\left( \nabla_{\mu}\partial^{\mu}\phi \right) - P_{,XX}\partial^{\mu}\phi \partial_{\alpha}\phi\left( \nabla_{\mu}\partial^{\alpha} \phi \right)   \nonumber \\
+P_{,X\phi}\partial^{\mu}\phi \partial_{\mu}\phi = \mathcal{L}_m \textup{Q},
\end{gather}
where $P_{,\phi}\equiv \partial P /\partial \phi$, $\nabla_{\mu}$ is the covariant derivative and the coupling term $\textup{Q}$ can be penned as
\begin{eqnarray}
\label{coupling}
\textup{Q} &=& -\frac{{f}_{,\phi}}{{f}} - \frac{{f}_{,X}}{{f}}\left[ \nabla_{\mu}\partial^{\mu}\phi+\partial^{\mu}\phi\left( \frac{\nabla_{\mu}\mathcal{L}_m}{\mathcal{L}_m} -\frac{{f}_{,\phi}}{{f}}\, \partial_{\mu}\phi \right. \right. \nonumber \\
&& + \left.\left.\frac{{f}_{,X}}{{f}}\, \partial_{\alpha}\phi\nabla_{\mu}\partial^{\alpha}\phi \right)\right] - \frac{{f}_{,X\phi}}{{f}}\,\partial^{\mu}\phi\partial_{\mu}\phi \nonumber \\
&& + \frac{{f}_{,XX}}{{f}}\,\partial^{\mu}\phi\partial_{\alpha}\phi\left( \nabla_{\mu}\partial^{\alpha}\phi \right).
\end{eqnarray}

The Bianchi identities declare that the total energy momentum tensor is conserved,
\begin{equation}
\nabla_{\mu}G^{\mu}_{\nu}=0\quad\Rightarrow\quad \nabla_{\mu}\left( T^{(\phi)\,\mu}_{\quad\,\,\,\nu}+T^{(m)\,\mu}_{\quad\,\,\,\,\,\nu} \right)=0,
\end{equation}
however, each individual component is not. The conservation relations take the form:
\begin{eqnarray}
\nabla_{\mu} T^{(\phi)\,\mu}_{\quad\,\,\,\nu} &=& \mathcal{L}_m \textup{Q}\,\nabla_{\nu}\phi, \label{consphi} \\
\nabla_{\mu}T^{(m)\,\mu}_{\quad\,\,\,\,\,\nu} &=& -\mathcal{L}_m \textup{Q}\,\nabla_{\nu}\phi,\label{consm}
\end{eqnarray}
where $\textup{Q}$ is given by Eq.\eqref{coupling}.

We remark that all the equations derived so far are completely general, in the sense that no specific choice for the metric $g_{\mu\nu}$ or for the coupling $f$ was assumed.

A large number of interacting dark energy models in the literature follow to impose the coupling at the level of the conservation relations \cite{Baldi:2012kt,Amendola:2000uh,Olivares:2006jr,Chen:2008ca,Quercellini:2008vh,Szydlowski:2016epq,Arevalo:2011hh,Nunes:2000ka}, choosing a particular form for the term on the right-hand side of Eqs.~\eqref{consphi} and \eqref{consm}. Interacting models with noncanonical scalar fields have also been studied \cite{Gumjudpai:2005ry,Chiba:2014sda,Shahalam:2017fqt}, which may allow couplings to depend nonlinearly on $\dot{\phi}$ \cite{Das:2014yoa}. Here however, the coupling is imposed at the level of the action, by choosing a particular form for $f(\phi,X)$, and then, the conservation relations naturally emerge {\it a posteriori} from this choice. Similar studies were explored, with a canonical scalar field in \cite{Koivisto:2005nr}, and for a tachyon field in \cite{Farajollahi:2011jr}, for the case where $f\equiv f(\phi)$.

\subsection{Particular cases}
\label{specificcases}

Inspecting Eq.\eqref{action}, we easily note that by neglecting the coupling (taking $f=1$), we recover the standard $k$-essence models extensively studied in the literature \cite{ArmendarizPicon:2000ah,GonzalezDiaz:2003rf,Scherrer:2004au,Babichev:2007dw,Chimento:2003ta,dePutter:2007ny}. Moreover, in the case of a canonical scalar field,
\begin{equation}
\label{canonical}
P(\phi,X) = X-V(\phi),
\end{equation}
where $V(\phi)$ is the potential function, the theory reduces to a quintessence model describing the dark energy source with a standard, noninteracting, canonical scalar field \cite{Zlatev:1998tr}. Setting ${f}\equiv{f}(\phi)$ yields coupled quintessence models \cite{Amendola:1999er} that were explored in, for example, \cite{Koivisto:2005nr}.

Assuming that our matter component plays the role of a pressureless fluid, we may write its Lagrangian density as \cite{Minazzoli:2012md,Harko:2010zi},
\begin{equation}
\label{matterlagrangian}
\mathcal{L}_{m}=-\rho_m,
\end{equation}
which holds for the energy-momentum tensor,
\begin{equation}
\label{emmattertensor}
T^{(m)}_{\mu\nu} = \rho_m u_{\mu} u_{\nu},
\end{equation}
the form of a pressureless perfect fluid, with $u_{\mu}$ being the four-velocity vector. In such case, it is interesting to notice that, when $f=f(\phi)$, Eqs.\eqref{consphi} and \eqref{consm} take the form:
\begin{eqnarray}
\nabla_{\mu} T^{(\phi)} \,^{\mu}_{\nu} &=&  \frac{{f}_{,\phi}}{{f}}\rho_m \nabla_{\nu}\phi, \label{a}\\
\nabla_{\mu} T^{(m)} \,^{\mu}_{\nu} &=& - \frac{{f}_{,\phi}}{{f}}\rho_m \nabla_{\nu}\phi, \label{b}
\end{eqnarray}
where, in this case, $\textup{Q}=-f_{,\phi}/f$.
These relations naturally arise in scalar-tensor theories \cite{Pettorino:2008ez}, in the Einstein frame, where the conformal factor is related with the coupling function through
\begin{equation}
\tilde{g}_{\mu\nu} ={f}(\phi)^2g_{\mu\nu}.
\end{equation}
Specifying ${f}(\phi)=e^{C\phi}$, where $C$ is a constant, we restore coupled quintessence models \cite{Amendola:1999er,Amendola:2014kwa,PhysRevLett.64.123}, with $\textup{Q}=-C$.
The case where the coupling is a function of the field, $C=C(\phi)$, was explored in \cite{Amendola:2003wa}.

Equations \eqref{a} and \eqref{b} hold an equivalence to a scalar tensor theory only because we have assumed a pressureless fluid form, given by Eq.\eqref{emmattertensor}, in which case the matter Lagrangian, Eq.\eqref{matterlagrangian}, equals the trace of the energy momentum tensor \cite{Koivisto:2005nr},
\begin{equation}
\mathcal{L}_m=T^{(m)}=-\rho_m.
\end{equation}
Obviously, the equations would not be valid for a more general fluid.

\section{Specific solution}
\label{example}

\subsection{Kinetically coupled quintessence}
\label{ex1}

Now that we have presented the underlying formalism for our theory, we follow to particularize for a specific case. We will assume that the accelerated expansion is mediated by a canonical scalar field $\phi$, the quintessence field, with Lagrangian density given by Eq.\eqref{canonical}, kinetically coupled to a cold dark matter component (CDM), described by Eq.\eqref{matterlagrangian}. Here, we assume that the interacting matter species plays the role of dark matter, as couplings to other fields, such as radiation or baryons, are strongly constrained by solar system constraints \cite{Anderson:2017phb,Carroll:1998zi,Devi:2011zz,Bertolami:2013qaa,Faraoni:2008ke,Davis:2007id}. Because the specific form for the dark matter Lagrangian is still unknown, here we assume that it can be expressed through a pressureless fluid, Eq.~\eqref{matterlagrangian}. Without loss of generality we will work with units where $8\pi G=1$.

Regarding the line element, we will stand on a flat Friedmann-Lema\^itre-Robertson-Walker background Cosmology,
\begin{equation}
\label{metric}
ds^2 = -dt^2 +a(t)^2 \delta_{ij}dx^idx^j,
\end{equation}
where $a(t)$ is the scale factor of the Universe as a function of cosmic time $t$.

We will assume an exponential potential of the form
\begin{equation}
\label{potential}
V=V_0\,e^{-\lambda \phi},
\end{equation}
where $V_0$ is a constant with dimensions of mass$^4$ and $\lambda$ is a dimensionless constant, expressing the stiffness of the potential. Potentials of the form Eq.\eqref{potential} have shown to present viable cosmological scaling solutions \cite{Copeland:1997et,Guo:2003eu,Barreiro:1999zs,Amendola:2014kwa}.

For concreteness, we close the system by specifying the following form for the coupling function, depending solely on the kinetic term,
\begin{equation}
\label{couplingfunction}
{f}=X^{\alpha},
\end{equation}
where now, with Eq.\eqref{metric}, we have on the background $X=\frac{1}{2}\dot{\phi}^2$, and $\alpha$ is a constant that dictates the strength of the kinetic coupling. When $\alpha=0$ we recover the standard uncoupled quintessence model \cite{Bahamonde:2017ize}.

\setlength\abovecaptionskip{5pt}
\begin{table*}[htp]
\centering
\begin{tabular}{cccccccc} 
\hline\hline
{\bf Point}$\quad$ & $x_c$ & $y_c$ & $\Omega_{\phi}$ & $w_{\rm eff}$ & $w_{\phi}$ & Existence & Accel. \\ [1.7ex] 
\hline
{\bf (A)}$\quad$ & $0$ & $0$ & $0$ & $0$ & $-$ &  $\forall\,\alpha,\lambda$ & No  \\ 
{\bf (B}$^{\pm}${\bf )}$\quad$ & $\pm 1$ & $0$ & $1$ & $1$ & $1$ &  $\forall\,\alpha,\lambda$  & No \\ 
{\bf (C)}$\quad$ & $\frac{\lambda}{\sqrt{6}}$ & $\sqrt{1-\frac{\lambda ^2}{6}}$ & $1$ & $\quad\frac{\lambda^2}{3}-1\quad$ & $\frac{\lambda^2}{3}-1$ &  $0<\lambda^2\leqslant 6\,\,\,\,\,$  & $\lambda^2<2$ \\ 
{\bf (D}$^{\pm}${\bf )}$\quad$ & $\pm \sqrt{\frac{\alpha}{1+\alpha}}$ & $0$ & $\frac{\alpha}{1+\alpha}$ & $\frac{\alpha}{1+\alpha}$ & $1$ &  $\forall\,\alpha,\lambda$  & No \\ 
{\bf (E)}$\quad$ & $\quad\sqrt{\frac{3}{2}}\frac{1+2\alpha}{\lambda(1+\alpha)}\,\,\,$ & $\,\,\,\sqrt{\frac{3-2\alpha(1+\alpha)(\lambda^2-6)}{2\lambda^2 (1+\alpha)^2}}\,\,\,$ & $\,\,\,\frac{3-\alpha(1+\alpha)(\lambda^2 -12)}{\lambda^2 (1+\alpha)^2}\,\,\,$ & $\,\,\,\frac{\alpha}{1+\alpha}\,\,\,$ & $\,\,\,\frac{\lambda^2 \alpha (1+\alpha)}{3-\alpha(1+\alpha)(\lambda^2 - 12)}\,\,\,$ &  Eq.~\eqref{existE}  & No     \\ 
\vspace{-0.4cm}\\
\hline
\end{tabular}
\caption{Fixed points of the system \eqref{eq1}-\eqref{eq3}, respective relative energy densities, equation of state parameters, existence regions and whether they feature accelerated expansion.}
\label{tabela1}
\end{table*}
\setlength\abovecaptionskip{-5pt}

Through the $\nu=0$ component of Eqs.~\eqref{consphi} and \eqref{consm}, the components of our Universe are then found to evolve as:
\begin{eqnarray}
\ddot{\phi} + 3H\dot{\phi}+V_{,\phi} &=& \rho_m\,\textup{Q}, \label{motionphi}\\
\dot{\rho}_m + 3H \rho_m &=& -\dot{\phi}\, \rho_m \, \textup{Q},\label{continuity}
\end{eqnarray}
where $H\equiv \dot{a}/a$ is the Hubble rate, and from Eqs.\eqref{coupling}, \eqref{matterlagrangian}, \eqref{couplingfunction}, \eqref{motionphi} and \eqref{continuity} we find
\begin{equation}
\label{couplingS}
\textup{Q}=2\alpha \frac{3H\dot{\phi}-\lambda V}{\dot{\phi}^2 + 2\alpha\left( \rho_m +\dot{\phi}^2 \right)}.
\end{equation}
The evolution equations are subjected to the Friedmann constraint, the $00-$component of Eq.\eqref{EEs},
\begin{equation}
\label{eqfriedmann}
3 H^2 =  \rho_{\phi}+\rho_m,
\end{equation}
where the energy density and pressure of the field reads, as usual,
\begin{eqnarray}
\rho_{\phi}&=&\frac{1}{2}\dot{\phi}^2+V, \\
p_{\phi}&=&\frac{1}{2}\dot{\phi}^2-V.
\end{eqnarray}

The rate of change of the Hubble parameter is given by,
\begin{equation}
-2\dot{H}=\rho_m+\dot{\phi}^2.
\end{equation}

In order to examine the dynamics of the system, it is common practice in Cosmology to introduce the following set of dimensionless variables \cite{Copeland:1997et} (sometimes referred to as {\it expansion-normalized variables} \cite{wainwright2005dynamical}),
\begin{equation}
x^2 \equiv \frac{\dot{\phi}^2}{6 H^2},\quad y^2 \equiv \frac{V}{3 H^2},\quad z^2 \equiv \frac{\rho_m}{3 H^2},
\end{equation} 
and rewrite the equations Eqs.\eqref{motionphi} and \eqref{continuity} in the form of an autonomous system of first order differential equations,
\begin{eqnarray}
x'&=&-x\left( 3+\frac{H'}{H} \right) + \sqrt{\frac{3}{2}}\left( \lambda y^2 + \textup{Q}z^2 \right), \label{eq1} \\
y' &=& -y \left( \sqrt{\frac{3}{2}}\lambda x + \frac{H'}{H} \right), \label{eq2} \\
z' &=& -z\left( \frac{3}{2} + \frac{H'}{H} + \sqrt{\frac{3}{2}}x\,\textup{Q} \right), \label{eq3}
\end{eqnarray}
where a prime denotes derivative with respect to the number of $e$-folds, $N\equiv\ln a$, and
\begin{equation}
\frac{H'}{H}=-\frac{3}{2}\left( 1+ w_{\rm eff} \right),
\end{equation}
where
\begin{equation}
\label{eqweff}
w_{\rm eff}= x^2-y^2
\end{equation}
is the effective equation of state parameter.

The equation of state parameter for the quintessence field reads
\begin{equation}
\label{statephi}
w_{\phi}=\frac{p_{\phi}}{\rho_{\phi}}=\frac{x^2-y^2}{x^2+y^2}.
\end{equation}

The constraint Eq.~\eqref{eqfriedmann} can now be written as
\begin{equation}
\label{friedmann1}
1=x^2+y^2+z^2,
\end{equation}
and is used to replace $z$ in terms of $x$ and $y$, reducing the dimension of the dynamical system. The coupling parameter $\textup{Q}$, Eq.~\eqref{couplingS}, can also be rewritten in terms of the dynamical variables,
\begin{equation}
\label{coupl}
\textup{Q}= \alpha\,\frac{\sqrt{6}x-\lambda y^2}{x^2 + \alpha\left( 1+w_{\rm eff} \right)}.
\end{equation}
Consistently, when $\alpha=0$ the system reduces to the one studied in \cite{Amendola:2014kwa}.

Kindly note that, the coupling function $f\propto \dot{\phi}^{2\alpha}$, Eq.~\eqref{couplingfunction}, diverges if $(\alpha<0)\wedge (\dot{\phi}=0)$, and the coupling term, Eq.~\eqref{coupl}, also diverges for negative values of $\alpha$ along the values
\begin{equation}
y^2 = 1+x^2\left( 1+\frac{1}{\alpha} \right),
\end{equation}
giving rise to an ill-defined phase space. For this reason, henceforth we assume $\alpha$ to be non-negative, that is $\alpha\geqslant 0$.

\subsection{Fixed point analysis and invariant sets}
\label{tabelaP}

In this subsection we report the nature of the critical points of the linear autonomous system \eqref{eq1}-\eqref{eq3} in terms of the free parameters present: strength of the kinetic coupling $\alpha$ and the stiffness of the potential given by $\lambda$. For cosmological applications, we are mainly interested in three properties: existence, stability and whether they feature accelerated expansion. The existence can simply be analysed through the condition $0\leqslant\Omega_{\phi}=x^2+y^2 \leqslant 1$ and requiring that the critical points are real valued. The acceleration nature is easily found by requiring that the effective equation of state obeys $w_{\rm eff}= x^2 - y^2<-1/3$. By linearising the system and considering a small perturbation around each fixed point, its stability can be studied through the eigenvalues of the stability matrix. More explicitly, given a linearised autonomous system,
\begin{equation}
{\bf x}' = A {\bf x}, 
\end{equation}
where ${\bf x}=(x_1,...\,,x_n) \in \mathbb{R}^n$ and $A$ is an $n\times n$ matrix, the stability of the fixed points (where ${\bf x}'=0$) is determined by the eigenvalues of the matrix $A$ evaluated at the critical point (see, for example \cite{Bahamonde:2017ize}). The {\it Hartman-Grobman theorem} \cite{Perko:1991:DED:102732} guarantees that the linear approximation is effective in expressing the dynamical behavior of the non-linear system, near the critical points. A fixed point is said to be stable (unstable) if the eigenvalues are all negative (positive). If at least one eigenvalue is positive and one negative, the point is a saddle. We refer the reader to \cite{wainwright2005dynamical,Bahamonde:2017ize} for technical details and applications of dynamical systems to DE and modified gravity.

\begin{figure*}[htp]
\begin{center}
\includegraphics[width=1\textwidth]{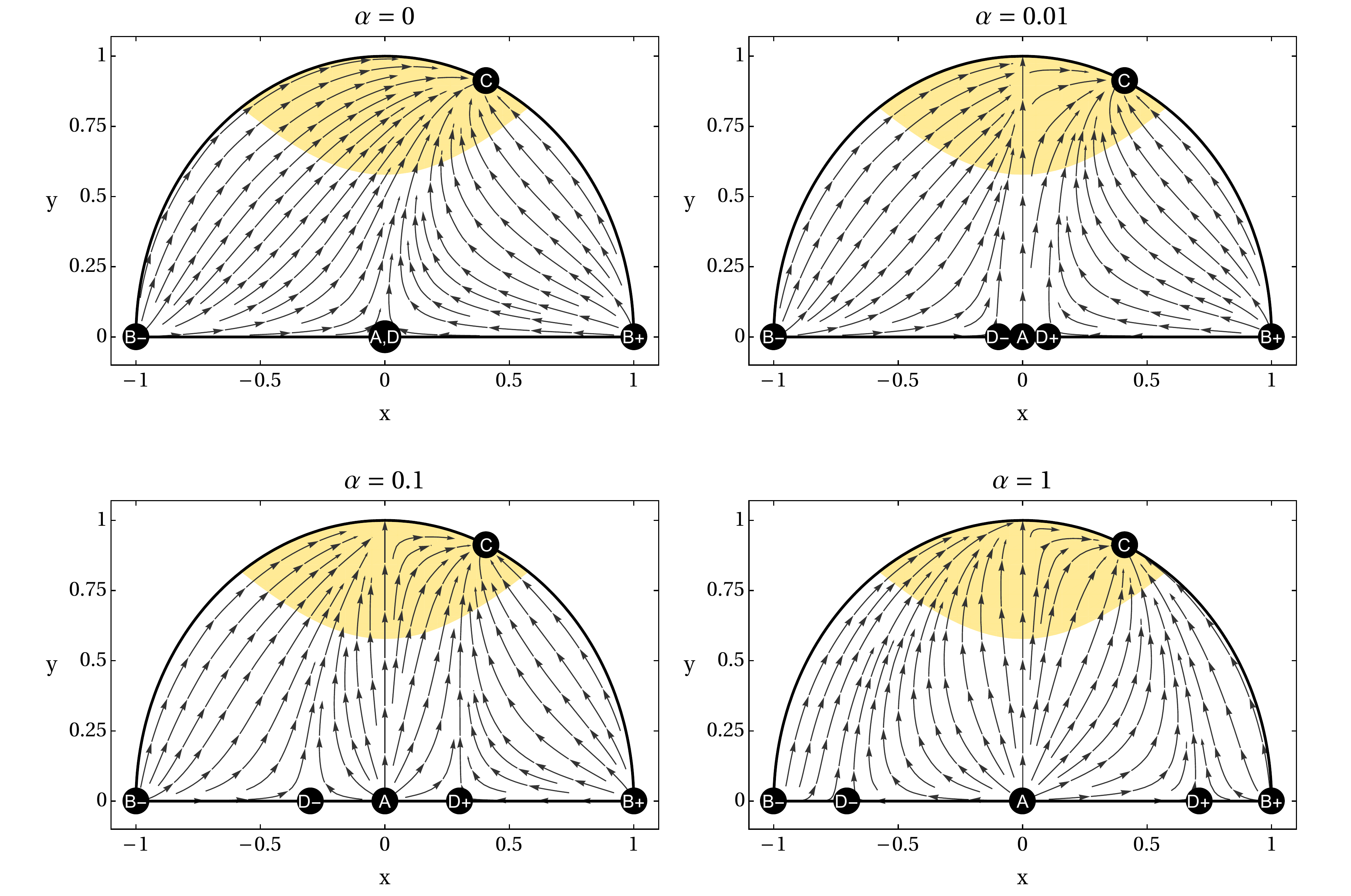}
\end{center}
\caption{\label{stream} Phase space of the system \eqref{eq1}-\eqref{eq3}, with $\lambda=1$, for different values of $\alpha$ (see respective title). The critical points are marked with black dots (with labels inside referenced to Table \ref{tabela1}). The shaded region corresponds to the area of the phase space for which the Universe is accelerating, i.e. where the condition $w_{\rm eff}  <-1/3$ is met.}
\end{figure*}

The system of Eqs.\eqref{eq1}-\eqref{eq3} is symmetric under the reflection $y\mapsto -y$ and time reversal $t\mapsto -t$. Therefore, we will only consider the upper half disk $y\geqslant0$ for our analysis. The dynamical system presents an additional symmetry, under the transformation $(x,\lambda)\mapsto (-x,-\lambda)$, meaning that the phase space is fully characterized if we only consider non-negative values of $\lambda$.

The fixed points of the system of Eqs.\eqref{eq1}-\eqref{eq3}, which are found through $(x',y')=(0,0)$, are presented in Table \ref{tabela1}.

\begin{itemize}
\item {\bf Point (A):} The origin of the phase space, $(x_c,y_c)=(0,0)$, corresponds to a matter dominated Universe, $\Omega_m=1$, and exists for all values of $\lambda$ and $\alpha$. This point does not generate accelerated expansion and is never stable. In the absence of coupling, $\alpha=0$, this fixed point has a saddle nature, attracting the trajectories along the $x$-axis and repelling them towards the $y$-axis. However, when $\alpha\neq 0$, this critical point acquires a repulsive nature. Thus, the presence of the kinetic coupling alters the dynamical nature of this critical point.
\item {\bf Point (B$^{\pm}${\bf ):}} These critical points correspond to scalar field kinetic dominated solutions, with $\Omega_{\phi}=x^2=1$, where the Universe is totally governed by the kinetic energy of the quintessence field. The equation of state describes a stiff fluid with $w_{\rm eff}=w_{\phi}=1$, and consequently, no acceleration is generated. These points exist for all values of $\lambda$ and $\alpha$ and are never stable. Critical point (B$^+$) is a saddle for $\lambda<\sqrt{6}$ ($\lambda>-\sqrt{6}$ for (B$^-$)) and a repeller otherwise.
\item {\bf Point (C):} This critical point corresponds to a scalar field dominated solution, $\Omega_{\phi}=1$. It exists for $\lambda^2 \leqslant 6$, and if the coupling is present, $\alpha\neq 0$, we also need to ensure that the parameter $\textup{Q}$, Eq.~\eqref{coupl}, is well defined. This translates into adding the condition $\lambda\neq 0$ for its existence when $\alpha\neq 0$. Acceleration is generated for $\lambda^2 < 2$. If one neglects the coupling, $\alpha=0$, this point is stable for $\lambda^2<3$. However, for $\alpha\neq 0$, the stability region can be enlarged into
\begin{equation}
\lambda^2 < 3 \frac{1+2\alpha}{1+\alpha}.
\end{equation}
This shows how the presence of the coupling can be used to widen the region in which (C) is an attractor. Fixed point (C) is a repeller when $\lambda^2>6$ and a saddle otherwise.

\item {\bf Point (D$^{\pm}${\bf ):}} These are novel critical points which emerge only if the kinetic coupling is present, and they exist for any values of $\alpha$ and $\lambda$. They correspond to a scaling solution, with $\Omega_{\phi}= \frac{\alpha}{1+\alpha}$, and coincide with the points (B$^{\pm}$) in the limit where $\alpha\rightarrow \infty$, and with (A) when $\alpha\rightarrow 0$. The effective equation of state parameter reads $w_{\rm eff}=\frac{\alpha}{1+\alpha}$, thus no acceleration is generated. Stability is achieved if
\begin{eqnarray}
\lambda > \sqrt{6+\frac{3}{2\alpha(1+\alpha)}},
\end{eqnarray}
for point (D$^+$),
\begin{eqnarray}
\lambda <- \sqrt{6+\frac{3}{2\alpha(1+\alpha)}},
\end{eqnarray}
for (D$^-$), and they acquire a saddle nature otherwise. As it will be shown in the next section, these critical points will play a major role on the cosmic evolution as they allow an early period of a scaling regime.
\item {\bf Point (E):} This critical point is also found for the uncoupled case, $\alpha=0$, \cite{Bahamonde:2017ize}, in which case, it exists for $\lambda^2\geqslant 3$. Here, however, it presents a generalized form with a dependence on $\alpha$. It corresponds to a scaling solution, with $\Omega_{\phi}=\frac{3-\alpha(1+\alpha)(\lambda^2 -12)}{\lambda^2 (1+\alpha)^2}$. In this model, when $\alpha\neq 0$, its existence is expressed through the region
\begin{eqnarray}
\label{existE}
6-\frac{3}{\alpha +1} \leqslant \lambda^2 \leqslant \frac{3}{2}\left( 4+\frac{1}{\alpha}-\frac{1}{1+\alpha} \right).
\end{eqnarray}
This critical point is always stable and it does not generate accelerated expansion.
\end{itemize}

\begin{figure}[]
\begin{center}
\includegraphics[width=.45\textwidth]{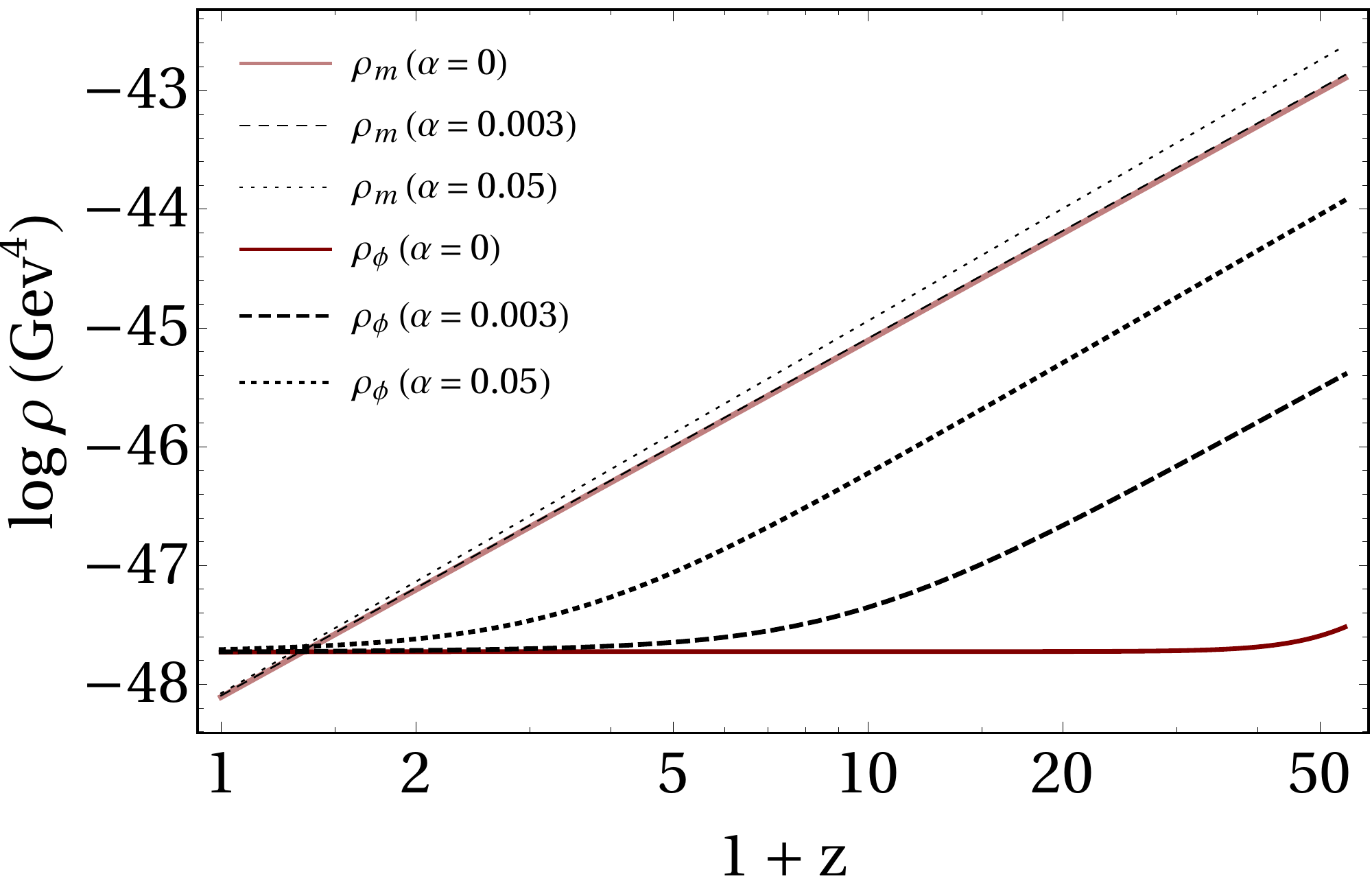}
\end{center}
\caption{\label{rhos}  Evolution of the energy densities for the field ($\rho_{\phi}$) and matter ($\rho_m$) for the solutions of Eqs.\eqref{eq1}-\eqref{eq3}, with $\lambda=0.2$ and different values of $\alpha$ (see legend).}
\end{figure}

Please note that, when $x=0$, Eq.~\eqref{eq1} reduces to $x'=0$. This means that,  when the kinetic coupling is present, $\alpha\neq 0$, the axis $x=0$ slices the phase space into two invariant sets (see \cite{wainwright2005dynamical}), $x<0$ and $x>0$: trajectories on these regions will always remain there. Therefore, assuming an expanding Universe, that is $H>0$, the field will always evolve in the same direction: the sign of $\dot{\phi}$ never changes in any physical orbit. The trajectory on the $x=0$ axis (also an invariant set itself) accounts for a $\Lambda$+CDM cosmology, with an uncoupled pressureless matter source and a cosmological constant with energy density $\rho_{\Lambda}=\rho_{\phi}=V= $ constant. Therefore, any trajectory with initial condition $x_i=0$, will inevitably mimic a $\Lambda$CDM cosmology, describing a transition from a matter dominated epoch, $(x,y)\rightarrow (0,0)$, into a potential totally dominated Universe, $(x,y)\rightarrow (0,1)$. This final state, mimics a cosmological constant governed solution, with $w_{\phi}=w_{\rm eff}=-1$, and $\Omega_{\phi}=y^2=1$.

There is an additional effect due to the slicing of the phase space. As stated above, when the kinetic coupling is present, the sign of $x$ does not change in any physical orbit. Therefore, if one wishes to reach the future attractor (C) with an initial condition satisfying $x_i<0$, we should also guarantee that $\lambda<0$ so that (C) lies on the $x<0$ plane of the phase space. The invariance of the system under the transformation $(x,\lambda)\mapsto (-x,-\lambda)$, guarantees that, by simultaneously changing the signs of $x_i$ and $\lambda$, the same cosmological evolutions are attained.

\subsection{Solutions and dynamical system analysis}
\label{dsanalysis}

The only fixed point that generates accelerated expansion and is stable is point (C). Therefore, we will rely on this point, a scalar field totally dominated solution (see Table \ref{tabela1}), to be the configuration towards which our Universe evolves into (a future attractor candidate). It is a well known, and widely studied, critical point that exists in scalar field models of dark energy and inflation \cite{Copeland:1997et,Copeland:2006wr}. Here however, points (E) and (D$^{\pm}$) are also stable for certain values of $\lambda$ and $\alpha$ whilst not featuring accelerated expansion. There is no risk of the trajectories falling into these points rather than (C), since the parametric window for $\lambda$ where (C) generates acceleration is outside the regions of existence of point (E) and stability of (D$^{\pm}$). Hence, in our simulations the critical point (E) is absent from the phase space and (D$^{\pm}$) are saddles.

\begin{figure}[t]
\begin{center}
\includegraphics[width=.45\textwidth]{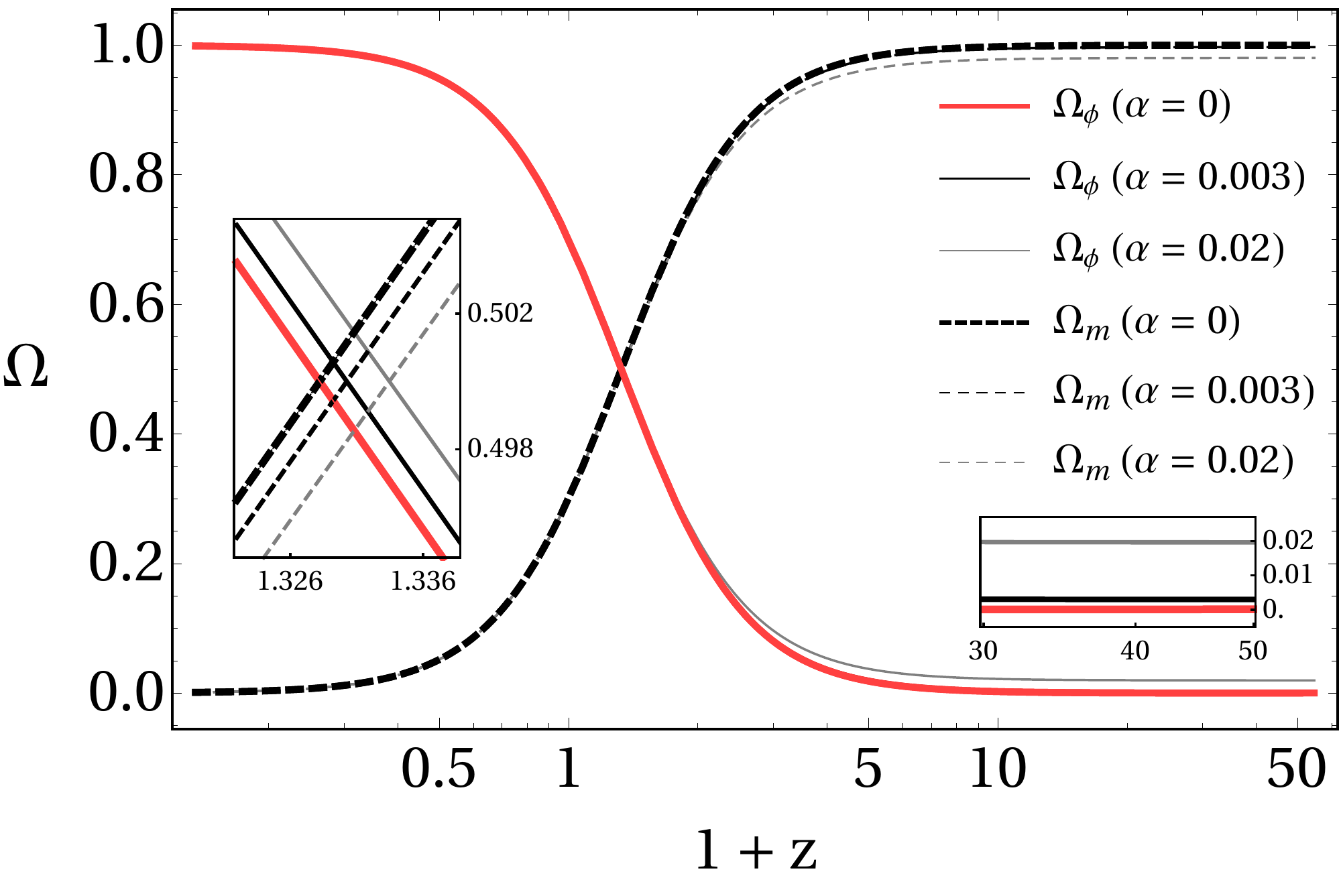}
\end{center}
\caption{\label{omegas} Relative energy densities $\Omega_{\phi}$ (solid) and $\Omega_m$ (dashed), for the solution of Eqs.\eqref{eq1}-\eqref{eq3}, with $\lambda=0.2$ and different values of $\alpha$ (see legend).}
\end{figure}

When we switch on the kinetic coupling, that is $\alpha \neq 0$,  we have the emergence of two new transient critical points, labeled as (D$^{\pm}$) in Table \ref{tabela1}. For small values of $\alpha$ they emerge near the origin, ${\rm  D}^{\pm} \xrightarrow{ \alpha \rightarrow\, 0\,\,} {\rm  A}$, and are shifted along the $x$-axis, until they coincide with the points (B$^{\pm}$) respectively, in the limit where the coupling is large, that is ${\rm  D}^{\pm} \xrightarrow{ \alpha \,\rightarrow\, \infty\,\,} {\rm  B^{\pm}}$. This effect can be better seen on Fig.\ref{stream}. These critical points are relevant for cosmological implications since they present a scaling behavior with $\Omega_{\phi} = \alpha/(1+\alpha)$ (see Table \ref{tabela1}). These scaling regimes are one enticing consequence of exploring models beyond the standard $\Lambda$CDM \cite{Albuquerque:2018ymr}, as they alleviate the cosmic coincidence problem by hiding the presence of the scalar field throughout an early period, where its energy density may be large, albeit small at present times. Note that in the current work, this new feature can only be achieved in the presence of the coupling, $\alpha\neq 0$, absent in the standard uncoupled model. As the critical point (D) is a saddle, we can naturally exit the scaling regime and proceed towards the accelerating attractor (C). To this end, we are interested in the heteroclinic orbits connecting (D)$\rightarrow$(C) (rigorously, orbits passing sufficiently close to them). They represent a transition from a matter dominated epoch to a scalar field totally dominated era.

To guarantee the scaling regime during the matter domination, we start the simulations near the critical point (D$^+$) with $x_i = \sqrt{\alpha/(1+\alpha)}$ (see Table \ref{tabela1}). The initial condition for $y$ is tuned with the parameters (always near the critical point (D) to ensure the scaling) such that at present times $\Omega_{\phi}^0\approx 0.7$ and $w_{\rm eff}\approx -0.7$, as it is suggested by cosmological observations \cite{Aghanim:2018eyx}. Finally, we start the simulations at $N_i=-4$ ($z\approx 54$), to ensure that we are in a matter dominated epoch.

Figure \ref{rhos} depicts the energy densities $\rho_{\phi}$ and $\rho_m$ for a particular set of solutions of Eqs.\eqref{eq1}-\eqref{eq3}, with $\lambda=0.2$ and different values of $\alpha$ (see legend). We clearly note that as the coupling emerges, $\alpha\neq 0$, a scaling regime arises, in contrast with the uncoupled quintessence model. During this epoch, the scalar field scales with matter as $\rho_{\phi}/\rho_m = \alpha$, eventually exiting this regime and becoming dominant as it evolves towards the attractor (C). Figure \ref{omegas} displays the evolution of the relative energy densities, $\Omega_{\phi}$ and $\Omega_m$, for various coupling parameters (see respective legend). We note that the transition from a matter to a DE dominated era happens earlier in the cosmic history for larger values of the coupling, as energy is being transferred from the dark matter component into the quintessence field. A similar behavior was found in \cite{Barros:2018efl} considering conformal couplings. Kindly note that the values of $\Omega_{\phi}$ for $\alpha=0.02$ in Fig.~\ref{omegas} suggest that the contribution of the scalar field energy at early times might be significant. This trend leads us to the following subject.

\begin{figure}[t]
\begin{center}
\includegraphics[width=.45\textwidth]{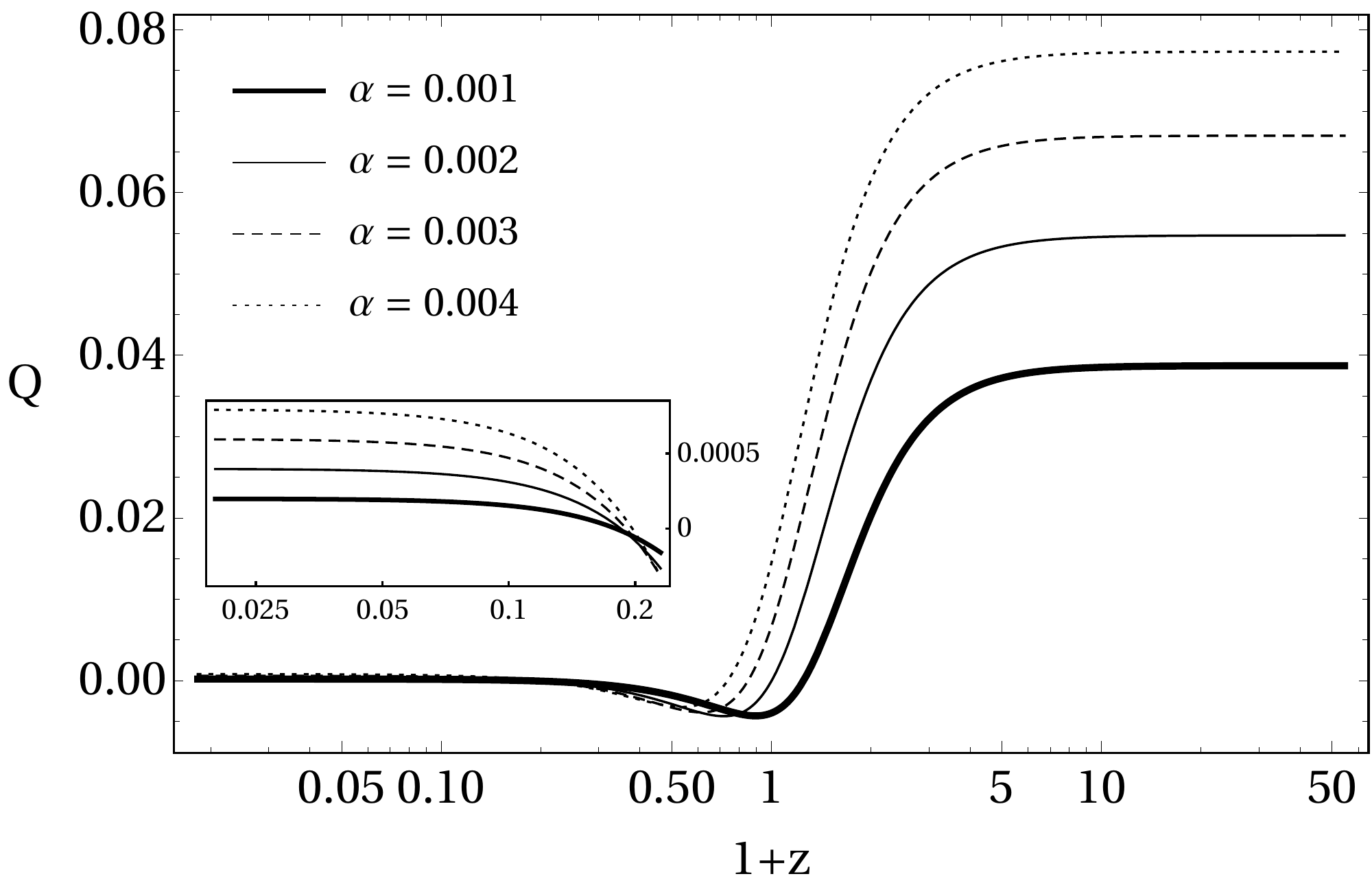}
\end{center}
\caption{\label{imcoup}  Interaction term $\textup{Q}$ for the solutions of Eqs.\eqref{eq1}-\eqref{eq3}, with $\lambda=0.2$, and different values of $\alpha$ (see legend).}
\end{figure}

It is known \cite{Wetterich:2004pv,Ade:2015rim} that the presence of a significant amount of early dark energy can substantially affect the position of the cosmic microwave background peaks. Therefore, this DE contribution at early times can be constrained by observations such as CMB lensing and small-scale measurements. From the Planck 2015 data \cite{Ade:2015rim} an upper bound of $\Omega_{\phi}<0.0036$ was found at $95$\% confidence level (for Planck TT,TE,EE+lowP+BSH). Regarding the present work, it is possible to use this constraint to bound the parameter $\alpha$, since during the scaling regime the energy density of the scalar field depends solely on this coupling parameter, $\Omega_{\phi}=\alpha/(1+\alpha)$ (see Table \ref{tabela1}). Thereupon, we find $\alpha < 0.0036$. This guarantees that the dark energy signatures are negligible during matter domination.  Note, however, that models with early dark energy may present a larger expansion rate at early times \cite{Karwal:2016vyq} and have been used to alleviate the $H_0$ tension \cite{Poulin:2018cxd}.

Following the scaling regime, the trajectories move towards the attractor (C) where an interesting demeanour occurs. Note that, even though the values of $x$ and $y$ at (C) are independent of $\alpha$ (see Table \ref{tabela1}), the value of the coupling parameter $\textup{Q}$ at the attractor is nonzero:
\begin{equation}
\textup{Q} = \frac{\alpha}{1+2\alpha}.
\end{equation}
We illustrate this behavior in Fig.\ref{imcoup}. This suggests that when the Universe is reaching the scalar field totally dominated era, the dark species carry on with their interaction. However, this synergy becomes negligible at the attractor, as $\rho_m \rightarrow 0$ (see Eqs.\eqref{motionphi} and \eqref{continuity}).

\begin{figure}[t]
\begin{center}
\includegraphics[width=.48\textwidth]{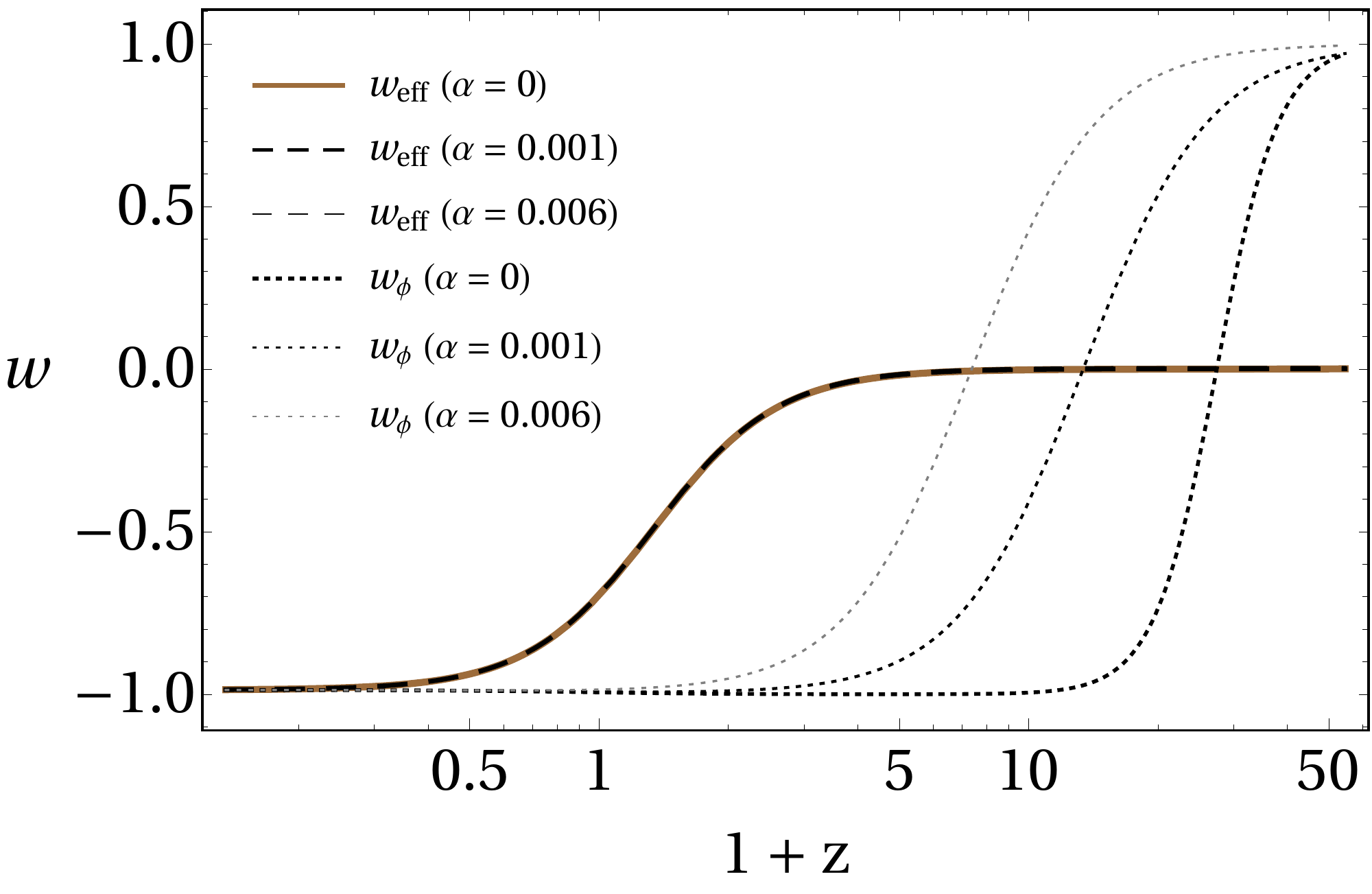}
\end{center}
\caption{\label{state}  Evolution of the effective equation of state parameter, $w_{\rm eff}$, and equation of state for the quintessence field, $w_{\phi}$, for the solution of Eqs.\eqref{eq1}-\eqref{eq3} with $\lambda=0.2$ and different coupling values (see legend).}
\end{figure}

In Fig.~\ref{state} we report the evolution of the effective equation of state $w_{\rm eff}$ and the equation of state for the quintessence field $w_{\phi}$. As shown earlier, when $\alpha = 0$ no scaling regime occurs and therefore the trajectory immediately begins to evolve towards the attractor (C). On the other hand, this departure is delayed in the presence of the interaction, due to the emergence of the scaling critical point (D$^+$). This effect becomes more noticeable with increasing $\alpha$. From Fig.~\ref{state} we also note that the previously described effect is dormant on the effective equation of state of the Universe, meaning that there are no significant changes on $w_{\rm eff}$ with increasing $\alpha$. This is ascribed to the fact that, for the coupling parameters chosen, we have ensured that the effects of DE are negligible during early times. When the transition to a dark energy era happens, the $\phi$ field will eventually dominate, and ultimately, accelerate the expansion (i.e. $w_{\rm eff}<-1/3$). Note that a cosmological constant behavior at the attractor, can only be achieved through $\lambda\rightarrow 0$, in which case $w_{\phi}\rightarrow -1$ (see Table \ref{tabela1}).

We finalize this section with a subtlety that should be addressed. The bounds found for the coupling parameter $\alpha$ may be alleviated if one shifts the early scaling regime towards latter times, where the constraints for $\Omega_{\phi}$ are not so stringent (see Fig.11 of \cite{Ade:2015rim}). In principle, one way of achieving this would be through choosing initial conditions further away from critical point (D) and closer to $(x_i,y_i)\approx (0,0)$, where the trajectories would enter the scaling regime later in the cosmic history. However, as this critical point is always a saddle in the present work, by choosing initial conditions far from it one would need to ensure that the solution would pass during a significant amount of time near point (D) - to guarantee the scaling - without being forthwith repelled towards the attractor. Nonetheless, an alternative to realize this effect might be considering a different choice for the potential, for instance a double exponential potential \cite{Barreiro:1999zs} where an early scaling is likewise present \cite{Amendola:2014kwa,Albuquerque:2018ymr}. In such case, by avoiding the Planck constraints, one procedure to test the model at lower redshifts could be through direct measurements of the Hubble rate, such as quasar distance measurements, from their X-ray and ultraviolet emission \citep{Risaliti:2018reu}. It was found deviations from $\Lambda$CDM at $4\sigma$ when taking into account data for $z>1.4$ (see Fig.3 of \citep{Risaliti:2018reu}). Furthermore, the future space-based interferometer eLisa \cite{Seoane:2013qna} is expected to be able to constrain early and interacting dark energy models at redshifts $1<z<8$ \cite{Caprini:2016qxs,Tamanini:2016zlh} from gravitational wave standard sirens.

\section{Conclusions}
\label{conclusions}

We explored a generalization of interacting dark energy models by allowing the kinetic term of a scalar field to couple to the matter sector at the level of the action. In section \ref{modell} we exposed the action and presented the modified field equations together with the individual conservation relations for the species.

In section \ref{example} we solved the equations for an FLRW Universe, considering a canonical scalar field interacting with a pressureless matter fluid (which is interpreted as a cold dark matter component) by specifying a particular form for the kinetic coupling.  The main feature found is the possibility of having cosmological solutions with an early scaling regime, where the effects of dark energy are negligible during the matter domination era, followed by a period of accelerated expansion, with a late time attractor. This early scaling regime becomes possible due to the emergence of two novel transient critical points, labelled as (D$^+$) and (D$^-$), when the kinetic coupling is present. These points represent a scaling solution with $\Omega_{\phi}=\frac{\alpha}{1+\alpha}$. These solutions are relevant for DE cosmologies since they can alleviate the cosmic coincidence problem, hiding the presence of the scalar field at the background level. During this regime, the energy density of the field may be large, despite its small value at present times. From Planck data, we were able to find an upper bound of $\alpha \leqslant 0.0036$ on the coupling parameter. One also finds that the presence of this interaction slices the phase space into two invariant sets, $x<0$ and $x>0$, leading to the fact that the sign of $\dot{\phi}$ is not allowed to change in any physical orbit. Thus, any trajectory with initial condition $\dot{\phi}_i=0$ freezes the field ($\phi=const.\,\,\forall N\geqslant N_i$) and inevitably reproduces a $\Lambda$CDM solution, evolving into a cosmological constant governed Universe with $\rho_{\phi}=\rho_{\Lambda}=V= const.$ and $w_{\phi}=w_{\rm eff}=-1$. Through the dynamical system analysis we have shown that the parametric region where the critical point (C) is an attractor can be enlarged by the presence of the coupling. Finally, the transition from a matter dominated epoch into a DE era happens earlier in the cosmic history when the coupling is stronger.

We highlight that the main difference of this theory, in contrast with the standard coupled quintessence models hitherto, is the underlying theoretical motivation. Here, the kinetic term of the field is allowed to couple to the matter fluids {\it a priori} in the action. As an aftermath of this assumption we reach a generalized form for the Bianchi identities, presented through the conservation relations for the species, Eqs.\eqref{consphi} and \eqref{consm}. As we have shown, this model encapsulates a plethora of existing models of dark energy as specific choices for the functions within the theory.

\acknowledgments

The author thanks Noemi Frusciante, Sante Carloni, Elsa Teixeira, Tiago Barreiro and Nelson Nunes for reading the manuscript and invaluable comments. The author would also like to thank the referee for the careful reading and useful references which have improved this current work. B. J. B. was supported by the grant PD/BD/128018/2016 from Funda\c{c}\~ao para a Ci\^encia e Tecnologia.

\bibliography{bib1}

\end{document}